\pdfoutput=1

\documentclass[twocolumn,twoside,nofootinbib,showpacs,prd,aps,tightenlines,10pt]{revtex4-1}

\usepackage{amsmath,braket,slashed}
\usepackage{graphicx}
\usepackage[hyperfootnotes=false,bookmarks=false]{hyperref}

\newcommand{\eq}[1]{Eq.~\eqref{eq:#1}}
\newcommand{\eqs}[2]{Eqs.~\eqref{eq:#1} and \eqref{eq:#2}}

\newcommand{\fig}[1]{Fig.~\ref{fig:#1}}

\def\nn{\nonumber \\ }

\def\df{{\rm d}}
\def\rd{{\rm d}}

\def\bn{\overline n}

\def\bnslash{\slashed{\bn}}

\def\bq{\overline q}

\def\al{\alpha}

\def\de{\delta}
\def\De{\Delta}

\def\si{\sigma}
\def\Pdef{\mathcal{P}}

\newcommand{\sieff}{\si_\mathrm{eff}}

\newcommand{\lqcd}{\Lambda_\mathrm{QCD}}

\allowdisplaybreaks[2]

\begin{document}

\title{What is Double Parton Scattering?}

\author{Aneesh V.~Manohar}

\author{Wouter J.~Waalewijn}

\affiliation{Department of Physics, University of California at San Diego,
  La Jolla, CA 92093\vspace{4pt} }

\begin{abstract}
Processes such as double Drell-Yan and same-sign $WW$ production have contributions from double parton scattering, which are not well-defined because of a
$\delta^{(2)}(\mathbf{z}_\perp=0)$ singularity that is generated by QCD evolution. We study the single and double parton contributions to these processes, and  show how to handle the singularity using factorization  and operator renormalization. We compute the QCD evolution of double parton distribution functions (PDFs) due to mixing with single PDFs. The modified evolution of dPDFs at $\mathbf{z}_\perp=0$, including generalized dPDFs for the non-forward case, is given in the appendix.  We include a brief discussion of the experimental interpretation of dPDFs and how they can probe flavor, spin and color correlations of partons in hadrons.
\end{abstract}

\maketitle

%~~~~~~~~~~~~~~~~~~~~~~~~~~~~~~~~~~~~~~~~~~~~~~~~~~~~~~~~~~~~~~~~~~~~~~~~~~~~~~~
\section{Introduction.}
%~~~~~~~~~~~~~~~~~~~~~~~~~~~~~~~~~~~~~~~~~~~~~~~~~~~~~~~~~~~~~~~~~~~~~~~~~~~~~~~

\begin{figure}[b]
\centering
\includegraphics[width=4.5cm]{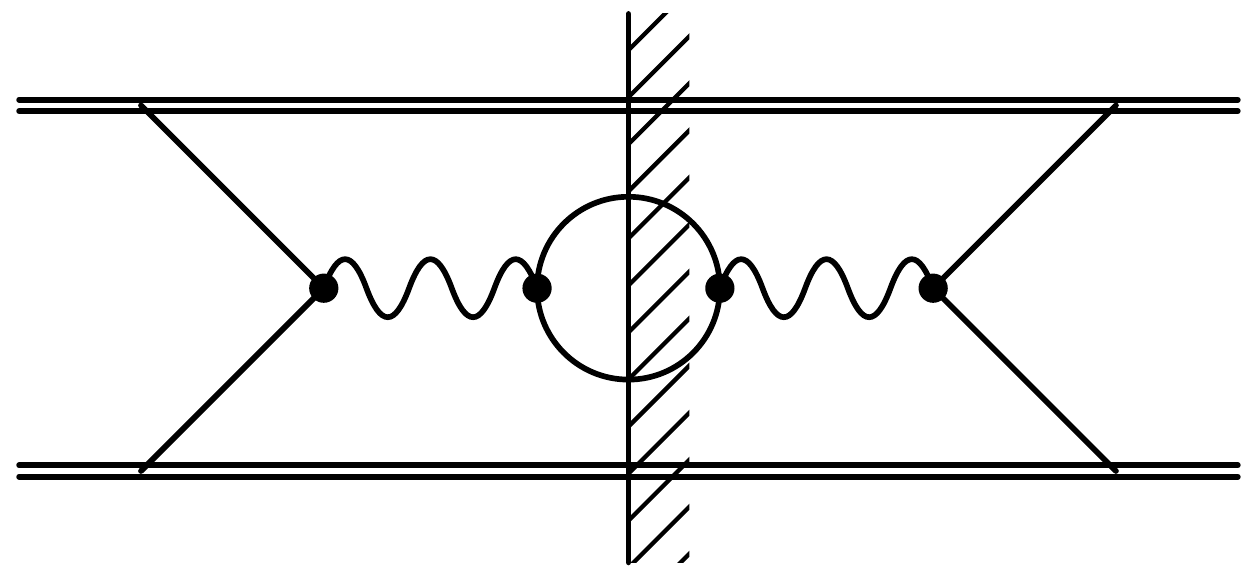}
\caption{Leading order diagram for single Drell-Yan. The cross section is shown as the cut of a forward-scattering amplitude.
The incoming hadrons are double lines, the gauge bosons are the wiggly lines, and the final state leptons are the cut loop. The hard interaction is given
by shrinking the gauge boson lines to a point.
\label{fig:DY}}
\end{figure}

Single parton scattering (SPS) processes such as Drell-Yan, $p_1 + p_2 \to \ell^+\ell^-$, shown in \fig{DY}, involve one parton in each hadron colliding via a hard interaction. Factorization allows one to write the cross section as 
%%%
\begin{eqnarray} \label{eq:1}
\frac{\df \si^\text{DY}}{\df x_1 \df x_2} = \frac{\hat \si_0}{x_1\, x_2} [f_q(x_1) f_{\bq}(x_2)+f_{\bq}(x_1) f_q(x_2)]
\,,\end{eqnarray}
%%%
at leading order. Here, $\hat \sigma_0 = 4\pi \alpha^2 Q_q^2/(3 N_c Q^2)$ is the short-distance $q \overline q \to \ell^+ \ell^-$ partonic cross section, $Q$ is the lepton-pair invariant mass which sets the hard scale, and $Q_q$ is the quark charge. The $f(x_i)$ are the (single) parton distribution functions (PDFs), and the momentum fractions $x_i$ are fixed by the invariant mass and rapidity of the lepton pair. The transverse momenta of the leptons are integrated over. If the transverse momenta of the leptons are measured, the expression in \eq{1} is modified and involves transverse-momentum-dependent PDFs instead.
\begin{figure}[b]
\centering
\includegraphics[width=5.5cm]{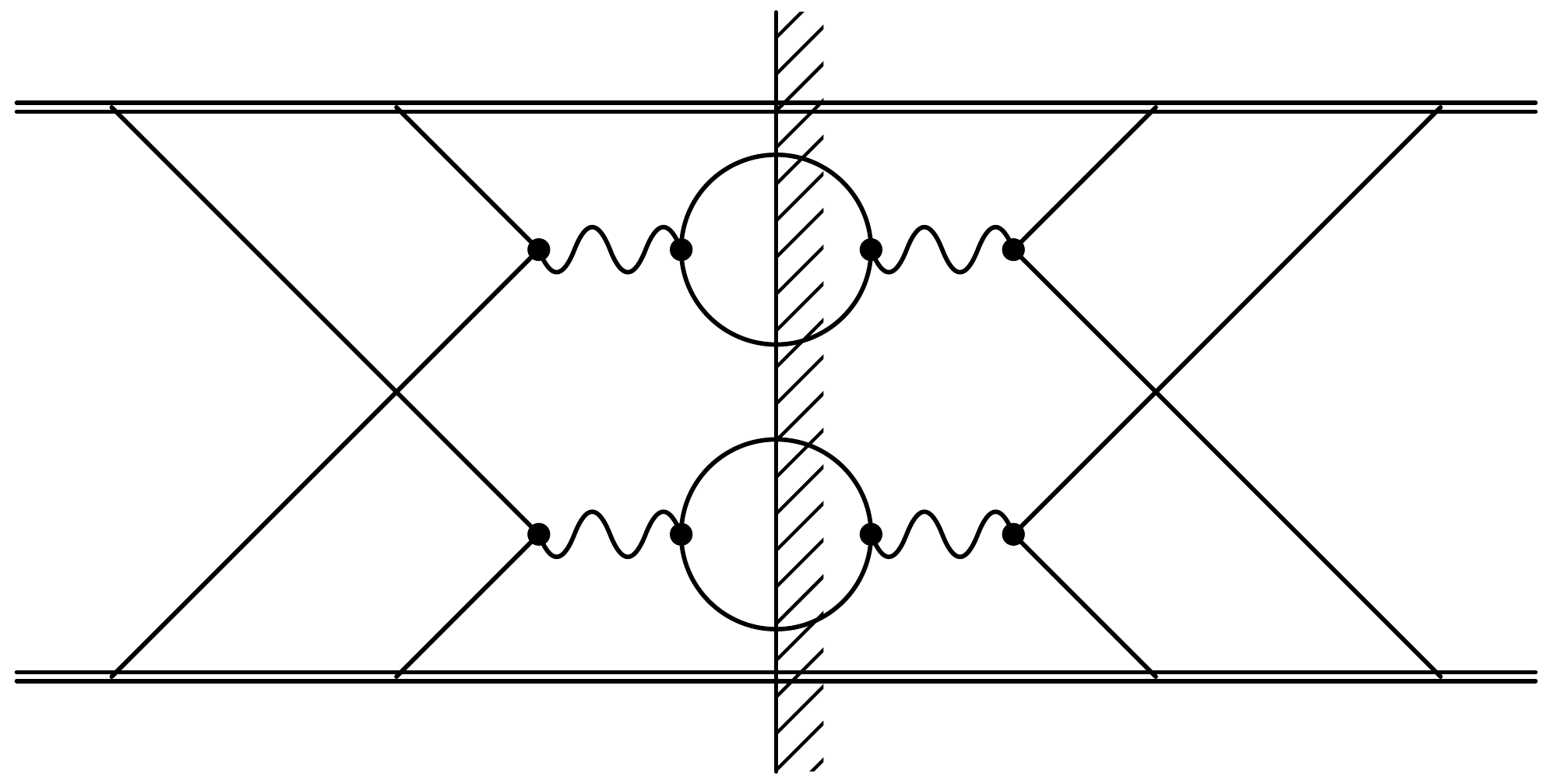}
\caption{Double parton scattering contribution to double Drell-Yan. The two hard interactions, given by shrinking the gauge boson lines to a point, are vertically separated in the figure by $\mathbf{z}_\perp$, which is not constrained by measurements.
\label{fig:DDY2}}
\end{figure}

One can also have processes where two partons in one hadron collide with two partons in the other hadron, which is known as double parton scattering (DPS). The leading-order DPS contribution to the double Drell-Yan cross section is shown in \fig{DDY2} and is (schematically) given by~\cite{Paver:1982yp}
%%%
\begin{eqnarray} \label{eq:si}
\frac{\df\si^\text{DPS}}{\df x_1 \df x_2 \df x_3 \df x_4} & \sim & \hat \si_0^2 \int\! \rd^2 \mathbf{z}_\perp\, F(x_1,x_2,\mathbf{z}_\perp) F(x_3,x_4,\mathbf{z}_\perp) 
\,.\nn \end{eqnarray}
%%%
The measured lepton-pair invariant masses and rapidities fix the momentum fractions $x_i$, and the transverse momenta are again integrated over. The hard double-parton scattering cross section $\hat \sigma_0^2$ is essentially the square of $\hat \sigma_0$ in \eq{1}. (The numerical factors can be found in Refs.~\cite{Mekhfi:1983az,Manohar:2012jr}.) $F(x_i,x_j,\mathbf{z}_\perp)$ are the double PDFs (dPDFs), which depend on the momentum fractions $x_i, x_j$ and the transverse separation $\mathbf{z}_\perp \sim 1/\lqcd$ of the two hard collisions. The transverse separation $\mathbf{z}_\perp$, or its Fourier-space analog $\mathbf{k}_\perp$, is not determined by the measurement and must be integrated over in \eq{si}, \emph{even} if one measures the transverse momenta of all the leptons. The dPDFs $F(x_i,x_j,\mathbf{z}_\perp)$ are of order $\lqcd^2$ and $\mathbf{z}_\perp$ is of order $1/\lqcd$. The $\mathbf{z}_\perp$ integral of two dPDFs is also of order $\lqcd^2$, so the DPS cross section is of order $\lqcd^2 \sigma_0^2$ and  $\lqcd^2/Q^2$ suppressed relative to SPS. 

There are additional terms that contribute to \eq{si} with spin and color correlations~\cite{Mekhfi:1983az,Mekhfi:1985dv} and interference effects~\cite{Diehl:2011tt,Diehl:2011yj}, which involve soft functions. Explicit expressions for their contribution can be found in e.g.~Ref.~\cite{Manohar:2012jr}, and they are briefly discussed in the appendix.
The color correlation and interference contributions are Sudakov suppressed at high energies~\cite{Mekhfi:1988kj,Manohar:2012jr}.

%~~~~~~~~~~~~~~~~~~~~~~~~~~~~~~~~~~~~~~~~~~~~~~~~~~~~~~~~~~~~~~~~~~~~~~~~~~~~~~~
\section{Mixing of single and double PDFs.}
%~~~~~~~~~~~~~~~~~~~~~~~~~~~~~~~~~~~~~~~~~~~~~~~~~~~~~~~~~~~~~~~~~~~~~~~~~~~~~~~

\begin{figure}
\centering
\includegraphics[width=5cm]{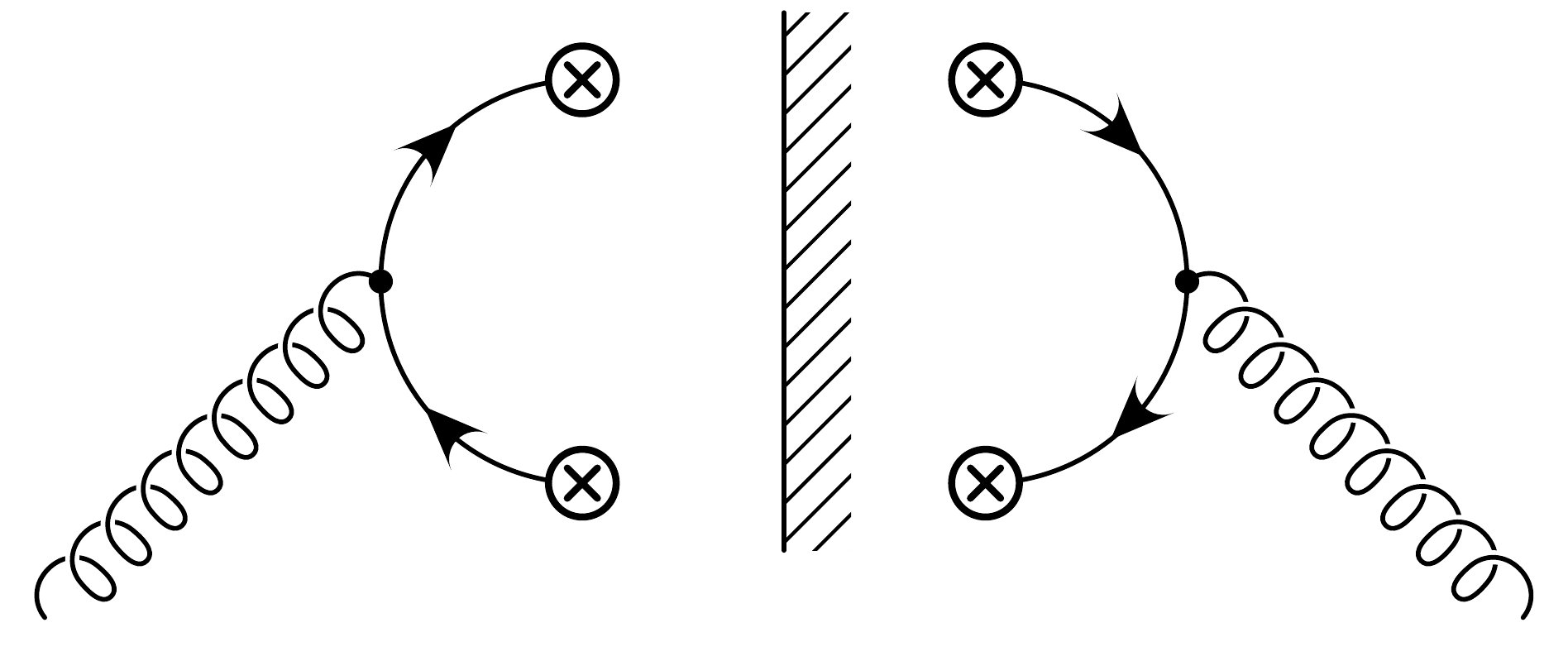}
\caption{Mixing of the gluon single PDF $f_g$ with the $q \overline q$ double PDF $F_{q\bq}$.}
\label{fig:mix}
\end{figure}

The intuitive description of DPS is somewhat misleading because there is also a contribution from SPS to double Drell-Yan, which mixes with DPS under the renormalization group evolution. For example, the graph shown in \fig{mix} leads to a mixing of the gluon PDF $f_g$ with the $q \overline q$ dPDF $F_{q\bq}$. This contributes to the renormalization group evolution of the color-summed dPDF $F^1_{q \overline q}$ as~\cite{Kirschner:1979im,Shelest:1982dg,Diehl:2011tt,Diehl:2011yj}
%%%
\begin{eqnarray} \label{eq:mix}
\mu \frac{\rd F_{q\bq}^1(x_1,x_2,\mathbf{z}_\perp)}{\rd \mu} &=& \frac{\alpha_s}{\pi}\, \de^{(2)}(\mathbf{z}_\perp)
P_{qg} \Big(\frac{x_1}{x_1\!+\!x_2}\Big) \frac{f_g(x_1\!+\!x_2)}{x_1+x_2},\nn
&&+\ldots
\,,\end{eqnarray}
%%%
in terms of the usual $g \to q$ splitting function $P_{qg}(x)$. The ``$\ldots$" denote terms given in Refs.~\cite{Kirschner:1979im,Shelest:1982dg,Diehl:2011tt,Diehl:2011yj,Manohar:2012jr} that do not involve mixing with single PDFs, and are not important for the present discussion. The $\de^{(2)}(\mathbf{z}_\perp)$ form of the mixing arises because the gluon splits into a $q \overline q$ pair at the same point.

We now discuss the modifications to \eq{mix} for the spin correlated, color correlated and interference double PDFs, as defined in Ref.~\cite{Manohar:2012jr}. Specifically,
the splitting function $P_{qg}$ is replaced by
%%%
\begin{eqnarray}
  P_{\De q g}(x) &=&  P_{q g}(x) = T_F [x^2 + (1-x)^2]\,, \nn
  P_{\de q g}(x) &=& -4T_F\, x(1-x)\,, \nn 
  P_{q g}^I(x) &=& P_{\De qg}^I(x) = -2T_F\, x(1-x) \,, \nn
  P_{\de q g}^I(x) &=& 2T_F [x^2+(1-x)^2]
\,,\end{eqnarray}
%%%
for $F_{\De q \De \bq}^1$, $F_{\de q \de \bq}^1$, $I_{q \bq}^1$, $I_{\De q \De \bq}^1$, $I_{\de q \de \bq}^1$, respectively. For the color-correlated dPDFs $F^T$ and $I^T$ the color factor $T_F$ is replaced by $(C_F-C_A/2)T_F$. The dPDFs $F_{q \delta \bq}$, $F_{\delta q \bq}$, $F_{\Delta q \delta \bq}$, $F_{\delta q \Delta \bq}$ and $F_{\delta q \delta \bq}^t$ and the corresponding interference dPDFs do not mix with the single PDFs.

As was pointed out in Refs.~\cite{Diehl:2011tt,Diehl:2011yj}, the $\de^{(2)}(\mathbf{z}_\perp)$ form of the mixing contribution in \eq{mix} is problematic. Under RG evolution $F_{q\bq}^1(x_1,x_2,\mathbf{z}_\perp)$ develops a $\de^{(2)}(\mathbf{z}_\perp)$ contribution, or equivalently, its $\perp$-Fourier transform $F_{q\bq}^1(x_1,x_2,\mathbf{k}_\perp)$ develops a $\mathbf{k}_\perp$-independent contribution. The integral of this term with the other dPDF in \eq{si} gives
%%%
\begin{eqnarray} \label{eq:one}
 &&\int\! \rd^2 \mathbf{z}_\perp\, \frac{\al_s}{\pi}\, \de^{(2)}(\mathbf{z}_\perp) f_{g}(x_1+x_2)\, F_{\bq q}^1(x_3,x_4,\mathbf{z}_\perp) \nn
 && \quad =  \frac{\al_s}{\pi} f_g(x_1+x_2) F_{\bq q}^1(x_3,x_4,\mathbf{z}_\perp=0)
  \,.\end{eqnarray}
%%%
However, there is also the convolution of the mixing contribution to both dPDFs with each other
%%%
\begin{eqnarray}  \label{eq:both}
 && \int\! \rd^2 \mathbf{z}_\perp\, f_g(x_1+x_2) \de^{(2)}(\mathbf{z}_\perp) f_g(x_3+x_4) \de^{(2)}(\mathbf{z}_\perp) \nn
 && \quad = f_g(x_1+x_2) f_g(x_3+x_4) \de^{(2)}(\mathbf{z}_\perp=0) = \,?
\,,\end{eqnarray}
%%%
which is singular. In momentum space, this singularity arises from the $\rd^2 \mathbf{k}_\perp$ integral of a constant. In this Letter we show how this problem is solved by a careful QCD factorization analysis of a physical cross section, such as double Drell-Yan. One hint is provided by the observation that the 
$\de^{(2)}(\mathbf{z}_\perp)$ contribution is when the $q \overline q$ pair are at the same $\mathbf{z}_\perp$, which  overlaps with the single parton scattering region.

%~~~~~~~~~~~~~~~~~~~~~~~~~~~~~~~~~~~~~~~~~~~~~~~~~~~~~~~~~~~~~~~~~~~~~~~~~~~~~~~
\section{Diagrammatic analysis.}
%~~~~~~~~~~~~~~~~~~~~~~~~~~~~~~~~~~~~~~~~~~~~~~~~~~~~~~~~~~~~~~~~~~~~~~~~~~~~~~~

%
%
%
\begin{figure}[t]
\centering
\includegraphics[width=3cm]{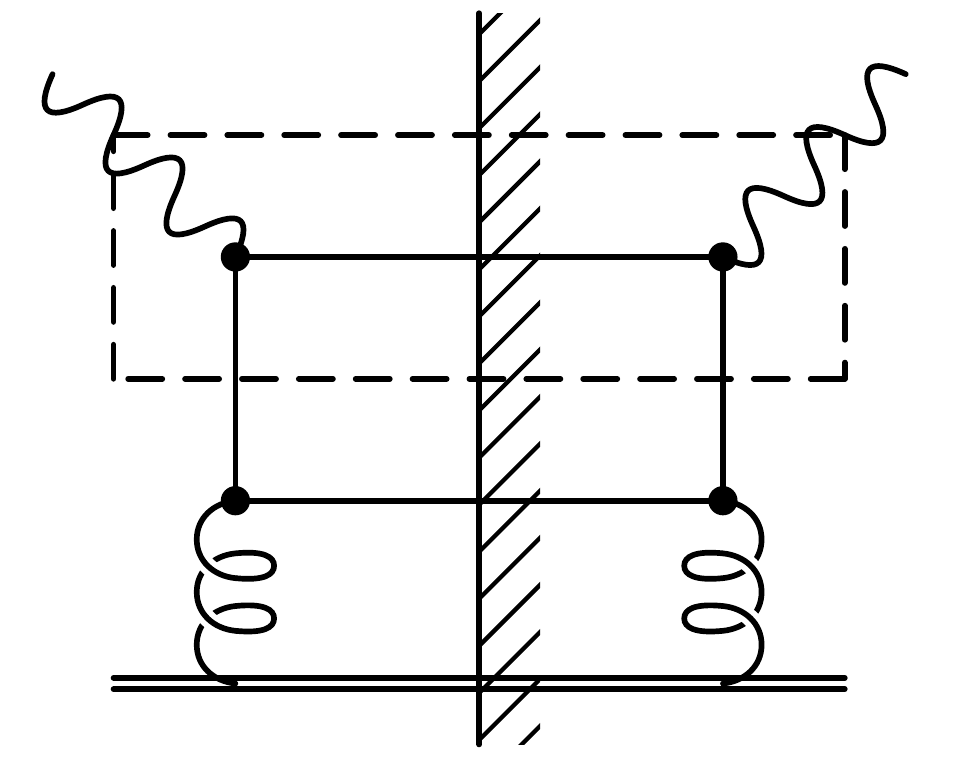}
\includegraphics[width=3cm]{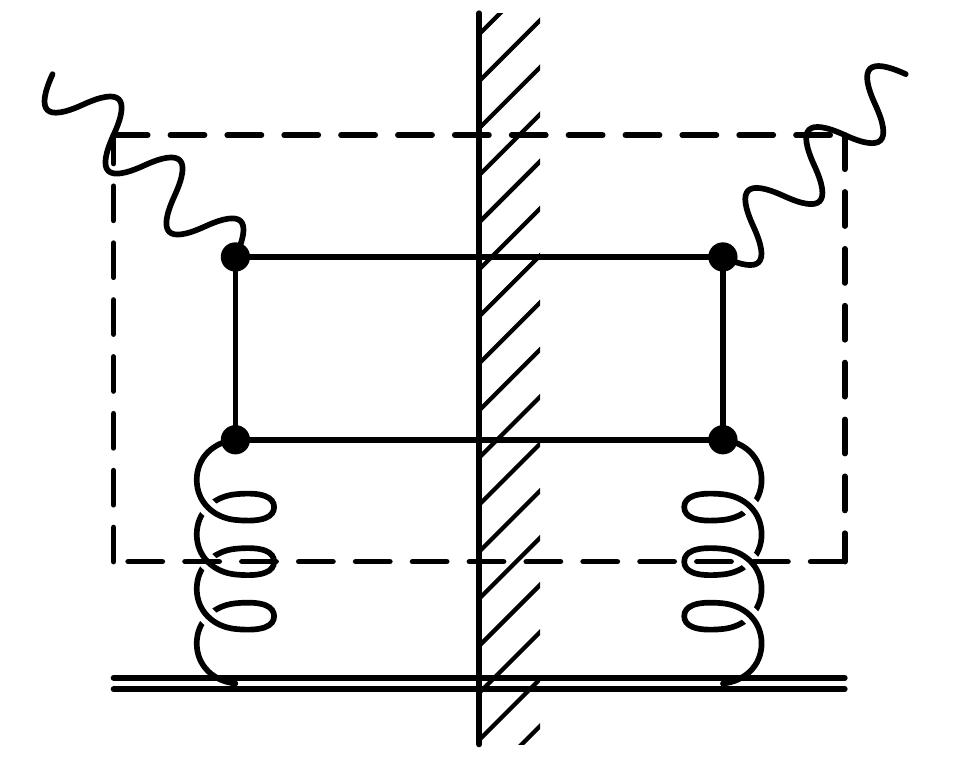}
\caption{Gluon contribution to deep inelastic scattering. The dashed area is the hard interaction, and is shrunk to a point. The wiggly lines are photons, and the springy lines are gluons.}
\label{fig:DIS}
\end{figure}
The Drell-Yan cross section in \eq{1} follows from factorizing the QCD graph for the cross section, shown in Fig.~\ref{fig:DY}.
To study (and define) DPS we look at a \emph{physical} process, such as double Drell-Yan. This receives both SPS and DPS contributions. Due to the mixing in \eq{mix}, the separation of the cross section into SPS and DPS contributions depends on the renormalization scheme and renormalization scale $\mu$. This is analogous to deep-inelastic scattering, where the cross section can be written in terms of quark and gluon distributions. The total is well-defined, but the split depends on the renormalization (or factorization) scheme and scale $\mu$. The same graph in Fig.~\ref{fig:DIS} gives the mixing of the quark and gluon PDFs (left figure) or the one-loop $\gamma g$ hard-scattering cross section (right figure). 

\begin{figure}
\centering
\includegraphics[width=5cm]{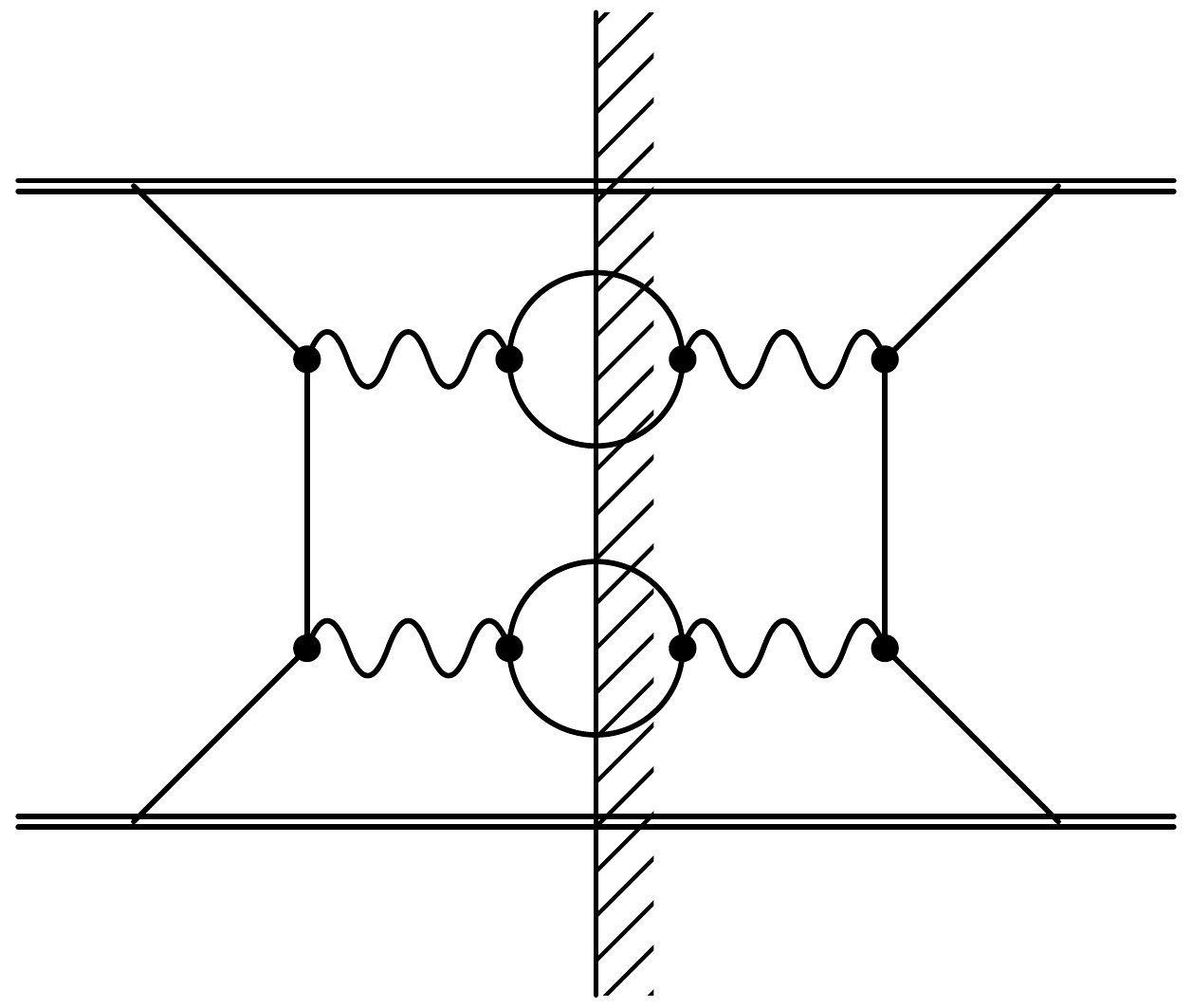}
\caption{Single parton scattering contribution to double Drell-Yan.}
\label{fig:DDY1}
\end{figure}

\begin{figure}
\centering
\includegraphics[width=6cm]{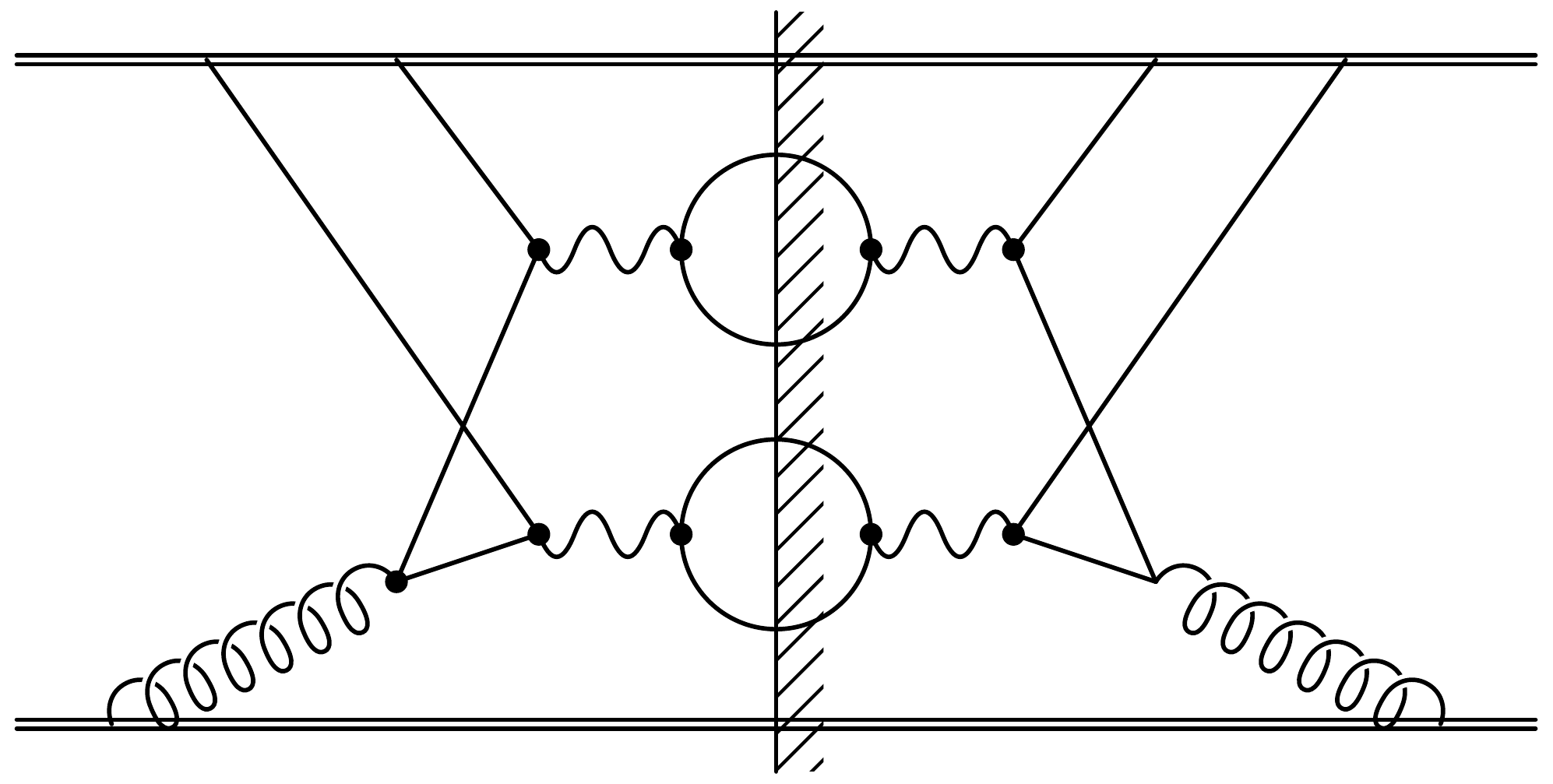}
\caption{Mixing contribution to double Drell-Yan.}
\label{fig:DDY3}
\end{figure}

\begin{figure}
\centering
\includegraphics[width=6cm]{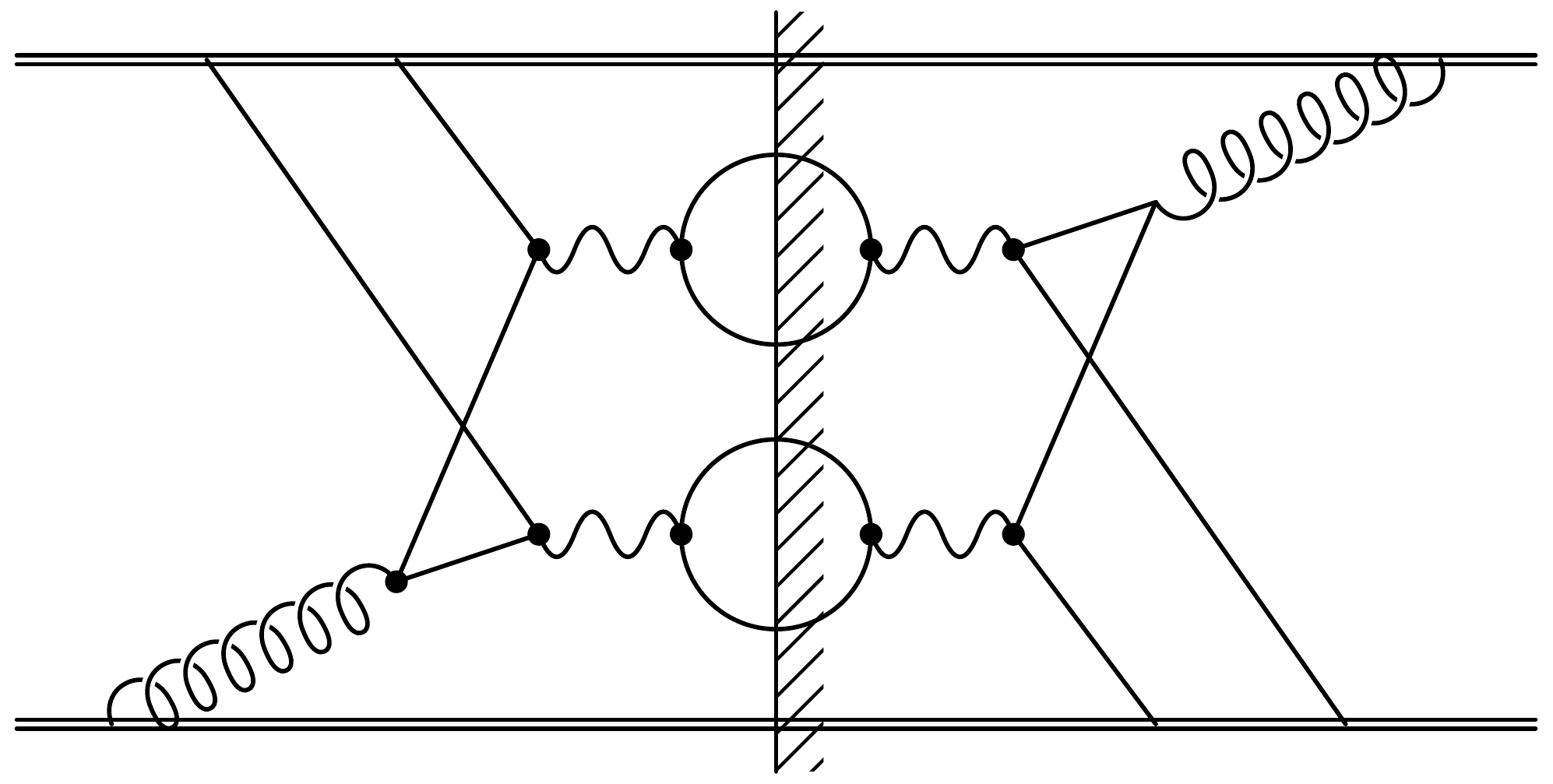}
\caption{Chirality suppressed mixing contribution to double Drell-Yan.}
\label{fig:DDY5}
\end{figure}

We now discuss the various contributions to the double Drell-Yan cross section, starting with the single parton scattering contribution in \fig{DDY1}.
The cross section is $[\alpha/(4\pi)]^2$ suppressed relative to \eq{1}, but is still leading twist. The leading double parton contribution is shown in \fig{DDY2} and leads to \eq{si}, and is $\lqcd^2/Q^2$ power suppressed. The mixing contribution is shown in Fig.~\ref{fig:DDY3}.
This graph has a short-distance contribution to double Drell-Yan of the form
%%%
\begin{eqnarray} \label{eq:6}
\sigma &\sim& \frac{\alpha_s}{4\pi} \sigma_0^2\  f_g(x_1+x_2) F_{\bq q}^1(x_3,x_4,\mathbf{z}_\perp=0)
\,,\end{eqnarray}
%%%
where the entire diagram is shrunk to a point. $F_{\bq q}^1(x_3,x_4,\mathbf{z}_\perp=0)$ is of order $\lqcd^2$, so this contribution is also $\lqcd^2/Q^2$ power suppressed.
This same diagram also has a contribution that corresponds to the mixing graph in \fig{mix} [in analogy with \fig{DIS}], and from \eq{one} we see that it is of the same size as \eq{6}. Finally, there is also the mixing graph shown in Fig.~\ref{fig:DDY5},
which leads to two twist-three PDFs, which are $q\overline q g$ matrix elements. However, these are chirality suppressed by light quark masses, and can therefore be neglected. 

\begin{figure}
\centering
\includegraphics[width=6cm]{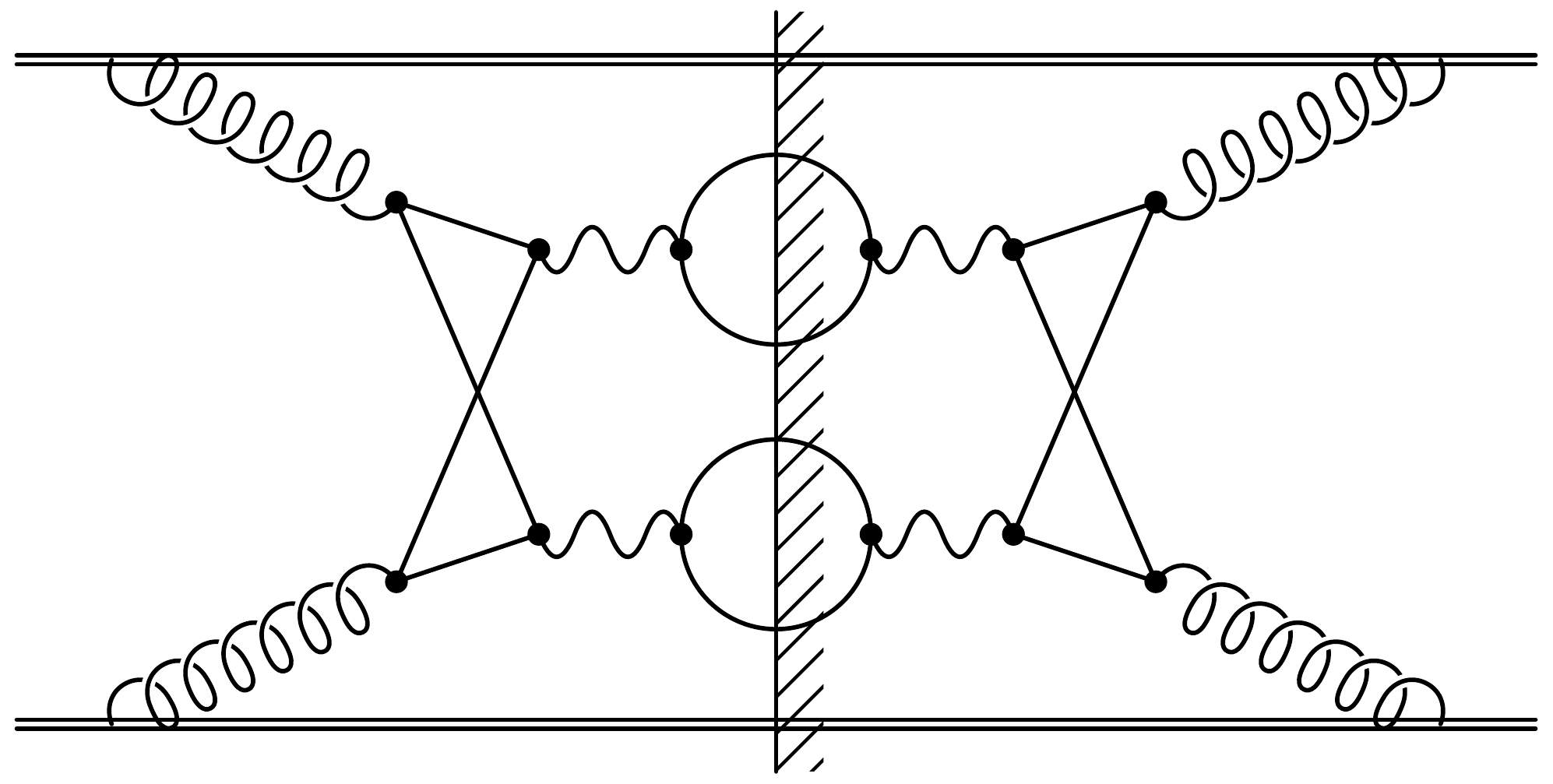}
\caption{Double mixing contribution to double Drell-Yan.}
\label{fig:DDY4}
\end{figure}

The problematic term in \eq{both} arises from the double mixing graph in Fig.~\ref{fig:DDY4}.
This graph is down by $[\alpha_s/(4\pi)]^2 [\alpha/(4\pi)]^2$ compared to single Drell-Yan, and is leading order in the twist expansion. It is of the same order in coupling constants as the $\left[\delta^{(2)}(\mathbf{z}_\perp)\right]^2$ term that arises from mixing. The part of the diagram that would produce \eq{both} is a quadratically divergent $\mathbf{k}_\perp$ integral. The two powers of $\mathbf{k}_\perp$ convert the leading-twist $f_g f_g$ gluon PDFs into the $\lqcd^2/Q^2$-suppressed dPDF contribution, since $|\mathbf{k}_\perp| \sim \lqcd$. However, this bound on $\mathbf{k}_\perp$ does not appear in perturbative calculations, and there is no physical scale of order $\lqcd$ that enters the graph. In dimensional regularization the quadratically divergent scaleless integral vanishes, so the contribution in \eq{both} is absent and there is no problem!

%~~~~~~~~~~~~~~~~~~~~~~~~~~~~~~~~~~~~~~~~~~~~~~~~~~~~~~~~~~~~~~~~~~~~~~~~~~~~~~~
\section{Operator renormalization.}
%~~~~~~~~~~~~~~~~~~~~~~~~~~~~~~~~~~~~~~~~~~~~~~~~~~~~~~~~~~~~~~~~~~~~~~~~~~~~~~~

The above discussion can be rephrased in terms of the renormalization of composite operators. The dPDF is
%%%
\begin{eqnarray}
\left[ F(\mathbf{z}_\perp)\right]
\,,\end{eqnarray}
%%%
keeping only the $\perp$ argument explicit. The square brackets emphasize that we are using the renormalized four-quark operator (defined explicitly in Refs.~\cite{Diehl:2011tt,Diehl:2011yj,Manohar:2012jr}) to get the finite dPDF. Fig.~\ref{fig:DDY2} gives a contribution
%%%
\begin{eqnarray}
\left[ \int\! \rd^2 \mathbf{z_\perp}\, F(\mathbf{z}_\perp)F(\mathbf{z}_\perp)\right]\,,
\end{eqnarray}
%%%
where the key point is that
%%%
\begin{eqnarray} \label{eq:9}
 \!\!\!\!\!\!\!\!\!\!\left[  \int\! \rd^2 \mathbf{z_\perp}\, F(\mathbf{z}_\perp)F(\mathbf{z}_\perp)\right] \not= 
\int\! \rd^2 \mathbf{z_\perp}\, \left[ F(\mathbf{z}_\perp) \right] \left[F(\mathbf{z}_\perp)\right] 
\,.
\end{eqnarray}
%%%
The integral over $\mathbf{z_\perp}$ includes the point $\mathbf{z_\perp}\!=\!0$, and one needs to perform an additional renormalization for products of operators at the same space-time point. This  is analogous to the well-known result that $\left[\phi^2(x)\right] \not = \left[\phi(x)\right]\left[\phi(x)\right]$ in an interacting theory.

\begin{figure}
\centering
\includegraphics[width=4cm]{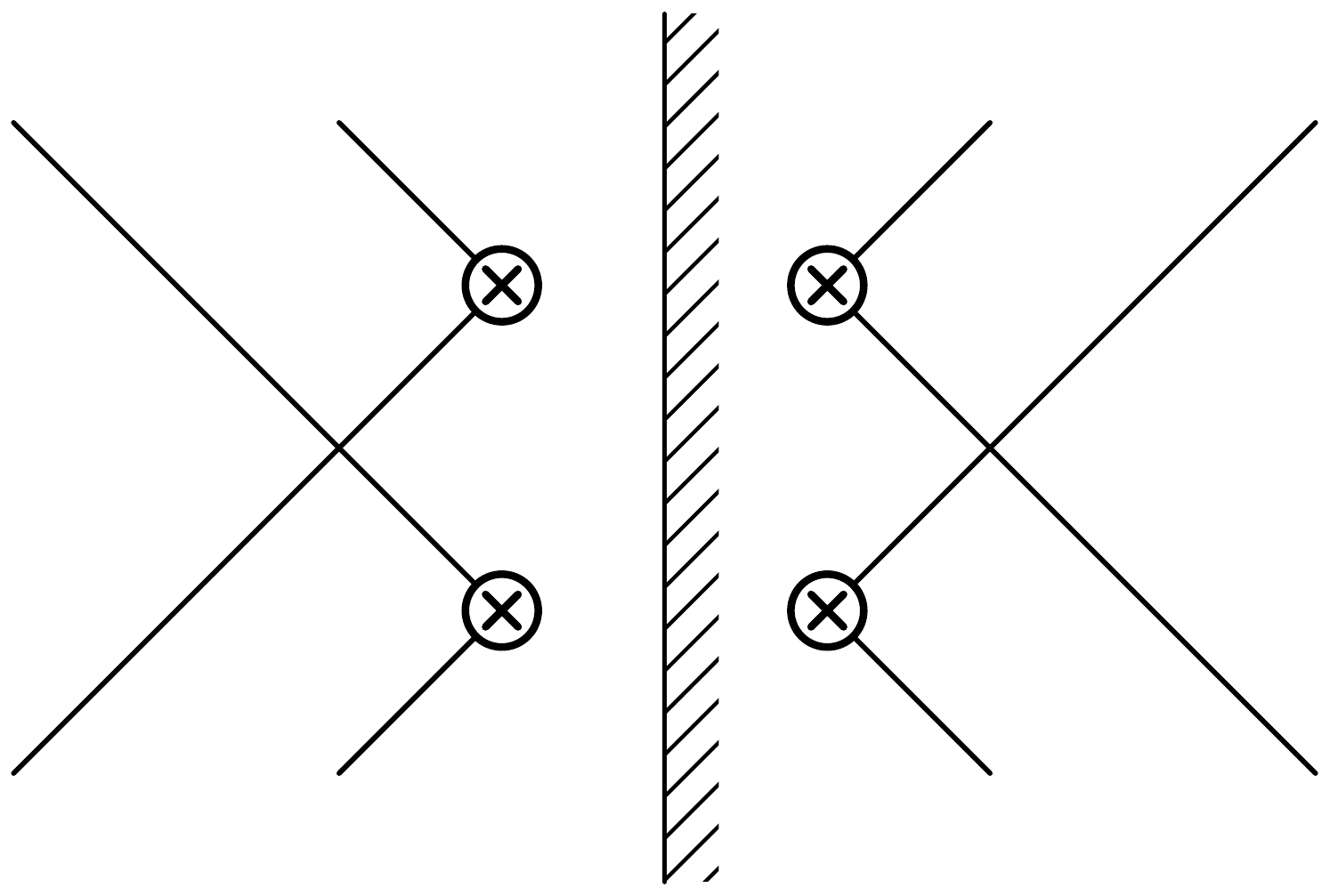}
\caption{The $\left[\int \rd^2 \mathbf{z_\perp}  F(\mathbf{z}_\perp)F(\mathbf{z}_\perp)\right]$ dPDF operator.
\label{fig:conv}}
\end{figure}

\begin{figure}
\centering
\includegraphics[width=4cm]{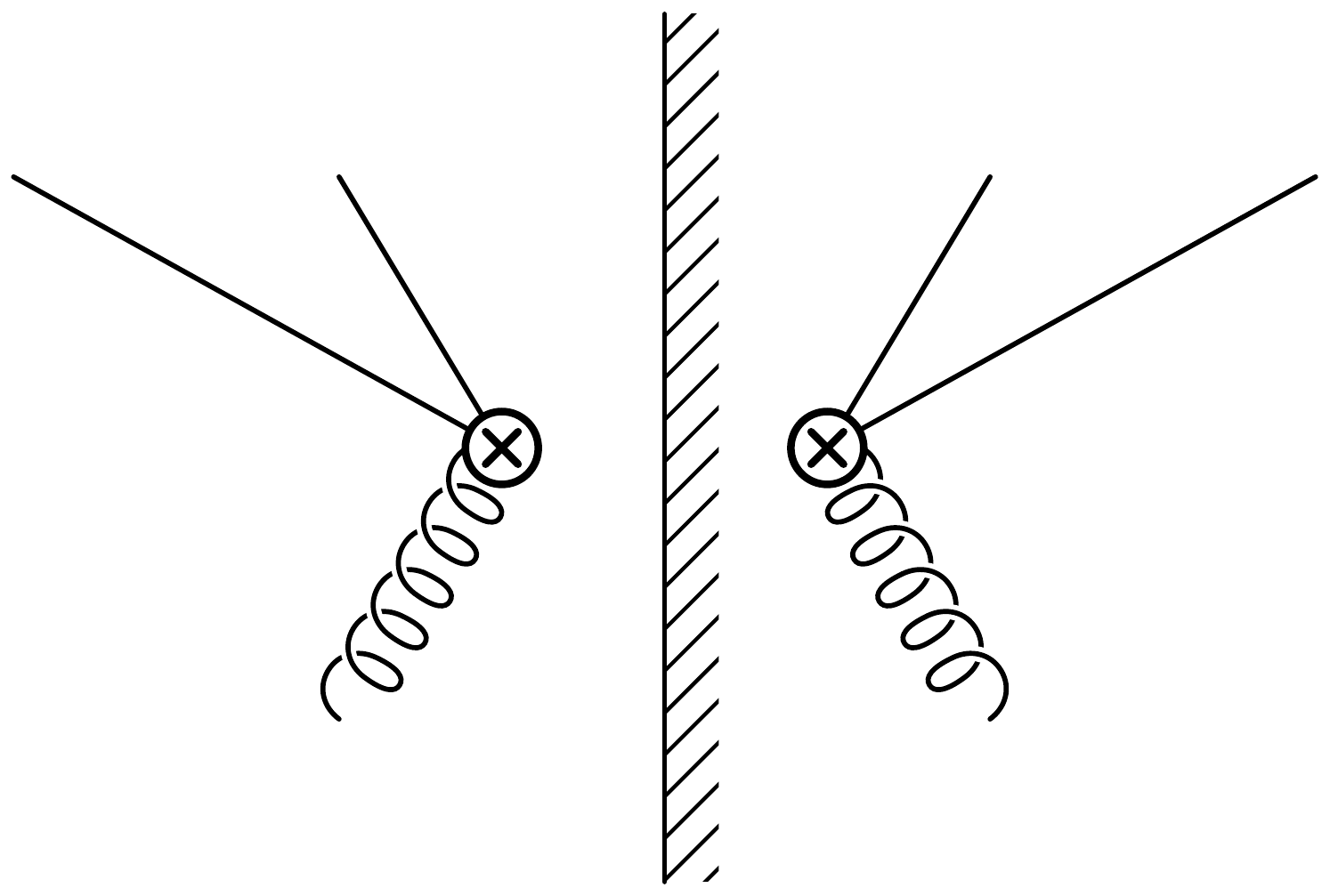}
\caption{The $\left[f_g F_{\bq q}^1(\mathbf{z}_\perp=0) \right]$ operator.
\label{fig:10}}
\end{figure}

We represent the dPDF operator in \eq{9} by Fig.~\ref{fig:conv}.
The mixing graph in Fig.~(\ref{fig:DDY3}) gives
%%%
\begin{eqnarray} \label{eq:10}
&& \mu \frac{\rd}{\rd \mu} \left[  \int\! \rd^2 \mathbf{z_\perp}\, F_{q\bq}^1(x_1,x_2,\mathbf{z}_\perp)F_{\bq q}^1(x_3,x_4,\mathbf{z}_\perp)\right] \\
 &&\quad = \frac{\alpha_s}{\pi}\,
P_{qg} \Big(\frac{x_1}{x_1\!+\!x_2}\Big) \left[\frac{f_g(x_1\!+\!x_2)}{x_1+x_2}F_{\bq q}^1(x_3,x_4,\mathbf{z}_\perp=0) \right]\nn
&&\qquad +  \frac{\alpha_s}{\pi}\,
P_{qg} \Big(\frac{x_3}{x_3\!+\!x_4}\Big) \left[F_{q\bq}^1(x_1,x_2,\mathbf{z}_\perp=0) \frac{f_g(x_3\!+\!x_4)}{x_3+x_4} \right]
,\nonumber \end{eqnarray}
%%%
where the right-hand side is the matrix element of the operator shown in Fig.~\ref{fig:10}.  Both sides of the equation are of order $\lqcd^2$.

\begin{figure}
\centering
\includegraphics[width=3.8cm]{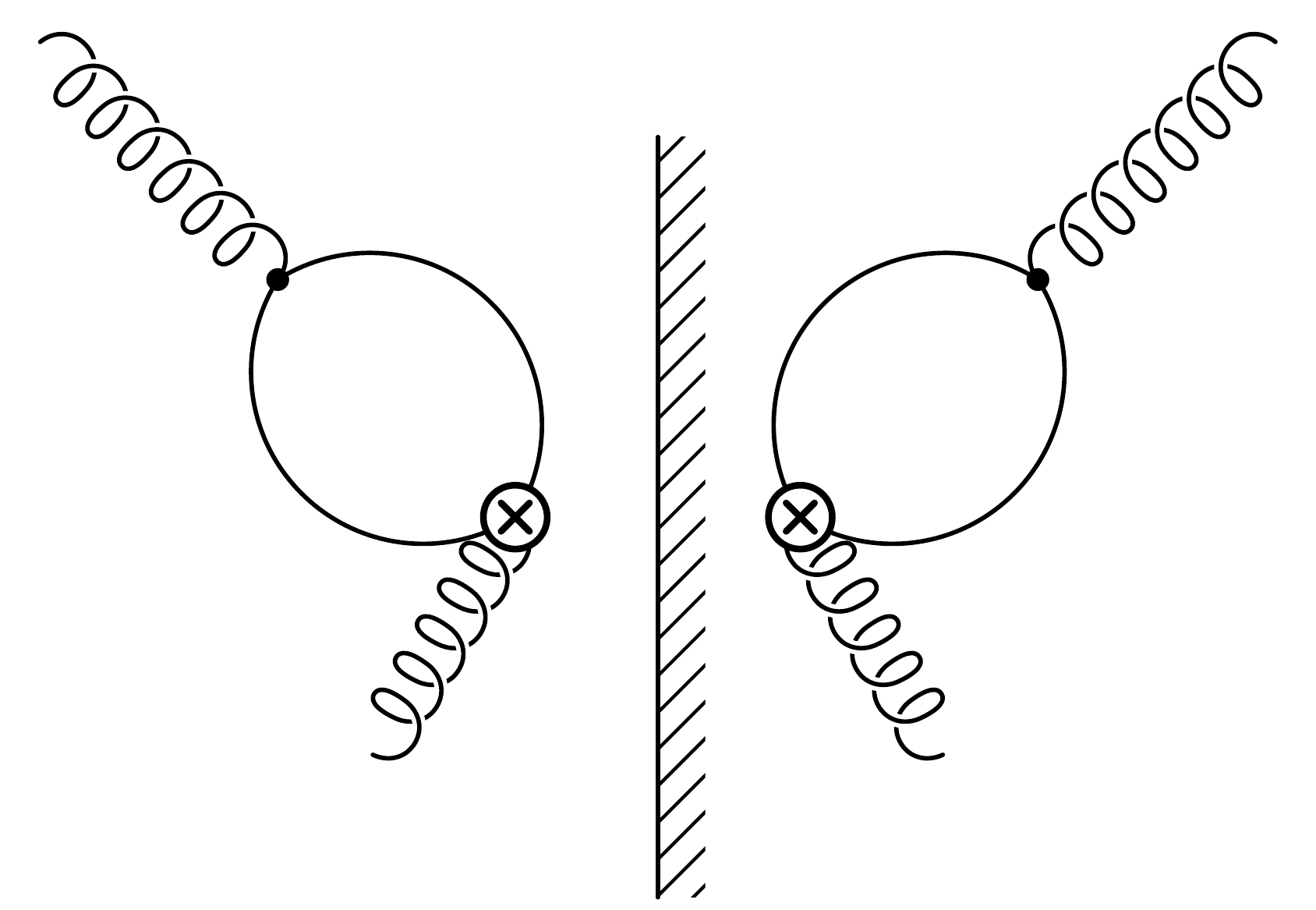}
\caption{Diagram for mixing the $\left[f_g\,F_{\bq q}^1(\mathbf{z}_\perp=0) \right]$ operator with gluon PDFs $f_g f_g$.}
\label{fig:11}
\end{figure}

The mixing of $\left[f_g  F_{\bq q}^1(\mathbf{z}_\perp=0) \right]$ into two single gluon PDFs shown in \fig{11} vanishes,
%%%
\begin{eqnarray}
 \mu \frac{\rd}{\rd \mu}  \left[f_g  F_{\bq q}^1(\mathbf{z}_\perp=0) \right]  =0\,.
 \label{eq:12}
 \end{eqnarray}
%%%
The left-hand side is order $\lqcd^2$, whereas the single gluon PDFs are order $\lqcd^0$. There is no order $\lqcd$ $\perp$ observable to compensate for the dimensions, so the two objects cannot mix. There is no longer any problem with a $\de^{(2)}(\mathbf{z}_\perp=0)$ as in \eq{both}.
Explicitly, the computation gives the integral
%%%
\begin{eqnarray}
\int \rd^{2-2\epsilon} \mathbf{k}_\perp &=& 0\,.
\label{zero}
\end{eqnarray}
%%%
This is the same integral that arises in \eq{both}. The difference is that Eq.~(\ref{zero}) is evaluated in fractional dimension using dimensional regularization, where it vanishes, whereas \eq{both} arises \emph{after} renormalization, and is evaluated in integer dimension, where it is singular.
The non-mixing contribution for the color-correlated and interference dPDFs are also modified at $\mathbf{z}_\perp=0$, as discussed in the appendix.

The above discussion shows that the singular quantity in \eq{both} does not enter the renormalization group evolution of the dPDF. Instead of Eq.~(\ref{eq:mix}) for the QCD evolution, one has the two equations (\ref{eq:10}) and (\ref{eq:12}).
Whereas we focus on the evolution of the dPDF, one may alternatively study the hard scattering (which correspond to the dPDF analog of the left and right panel in \fig{DIS} respectively). This second approach was taken in a recent paper by Gaunt and Stirling~\cite{Gaunt:2011xd}, in which they study the double parton scattering singularity in $gg \to Z Z$. They find that this is not logarithmically enhanced, indicating that $f_g f_g$ does not contribute to the leading-order cross section (through the RG evolution). They suggest that one should drop the $f_g f_g$ mixing term in \eq{both}, in agreement with our result. A related study was carried out in Ref.~\cite{Blok:2011bu}. They agree with Ref.~\cite{Gaunt:2011xd} that the $f_gf_g$ term is not logarithmically enhanced. However,
their cross section does contain an $f_g f_g$ mixing contribution because they use a cut-off rather than dimensional regularization.

%~~~~~~~~~~~~~~~~~~~~~~~~~~~~~~~~~~~~~~~~~~~~~~~~~~~~~~~~~~~~~~~~~~~~~~~~~~~~~~~
\section{Estimating the size of various contributions.}
%~~~~~~~~~~~~~~~~~~~~~~~~~~~~~~~~~~~~~~~~~~~~~~~~~~~~~~~~~~~~~~~~~~~~~~~~~~~~~~~

We thus have a well-defined factorization formula for a physical process such as double Drell-Yan. The factorized cross section has the following (schematic) form
%%%
\begin{eqnarray}
\sigma &\sim& \hat c_1 f_q f_{\bq} \!+\! \hat c_2 f_g f_g \!+\!  \hat c_3 \left[f_g F(\mathbf{z}_\perp=0) \right] \!+\!\hat c_3 \left[F(\mathbf{z}_\perp=0)f_g \right] \nn
&&+ \hat c_4 \left[\int \rd^2 \mathbf{z_\perp}  F(\mathbf{z}_\perp)F(\mathbf{z}_\perp)\right]\label{12}
\,,\end{eqnarray}
%%%
where $\hat c_i$ are the partonic cross sections for the hard scattering, and contains both single and double parton scattering contributions. In Eq.~(\ref{12}) we have left implicit the many possible different parton flavors, spin structures etc.~for the dPDFs. For double Drell-Yan, the $c_{1-4}$ terms contribute at order $[\alpha/(4\pi)]^2\,\sigma_0$, $[\alpha/(4\pi)]^2[\alpha_s/(4\pi)]^2\,\sigma_0$,
$[\alpha_s/(4\pi) ]\lqcd^2\sigma_0^2$ and $\lqcd^2 \sigma_0^2$, respectively to the cross section, where the coupling constants are evaluated at the hard scale $Q$.
\begin{figure}
\centering
\includegraphics[width=7cm]{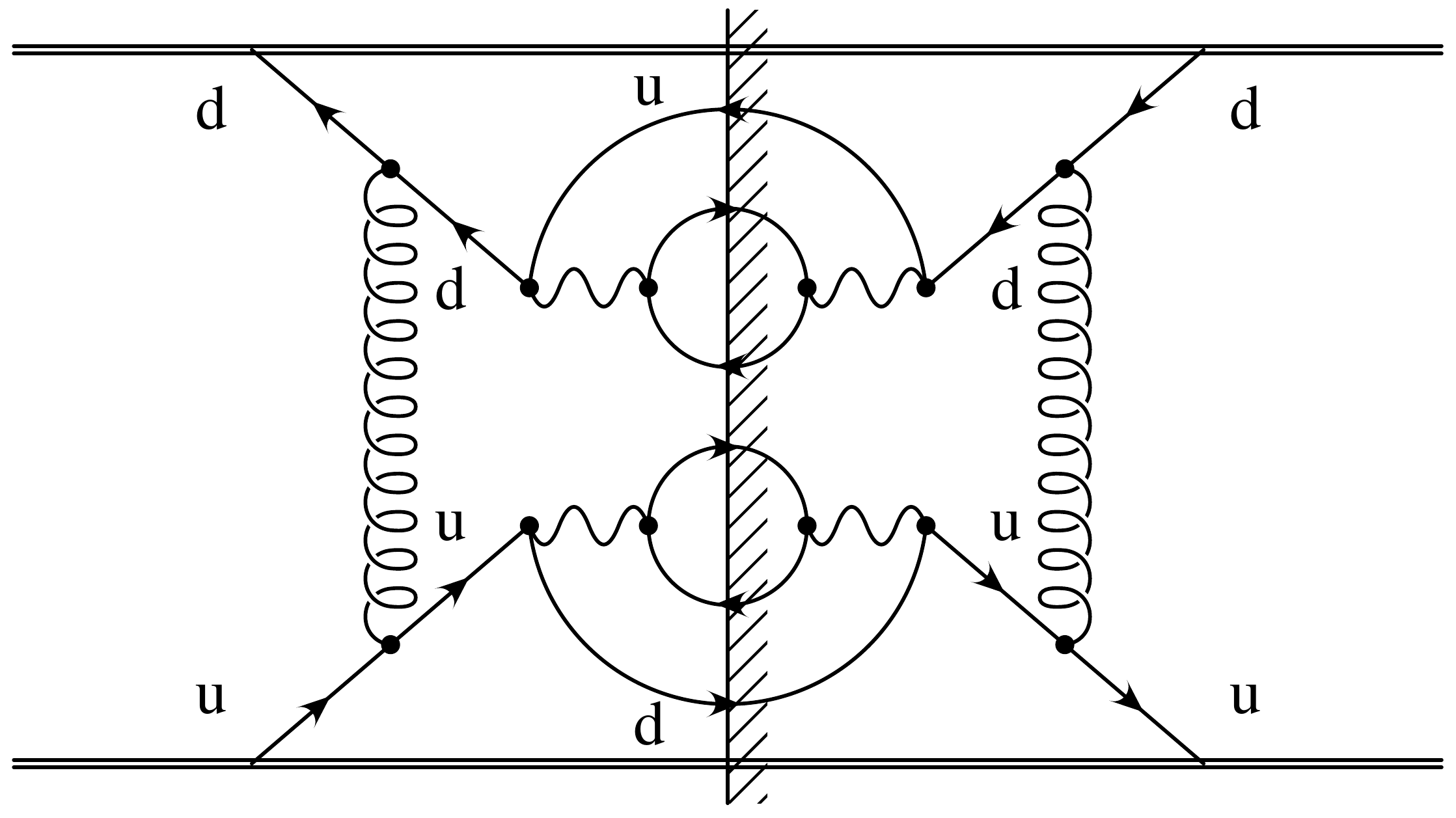}
\caption{Leading single PDF contribution to $W^+W^+$ production.
\label{fig:WW}}
\end{figure}
The $\hat c_{1,2}$ terms are leading power, whereas the $\hat c_{3,4}$ terms are $\lqcd^2/Q^2$ power suppressed.

Same-sign $WW$ production is a process in which the single PDF contribution is suppressed. The dPDF contribution is the same size as for double Drell-Yan, from Fig.~\ref{fig:DDY2}. However, the single PDF contribution requires two additional partons (jets) in the final state, as shown in Fig.~\ref{fig:WW}. The $\hat c_1$ contribution to the cross section is $\sim [\alpha/(4\pi \sin^2\theta_W)]^2 [\alpha_s/(4\pi)]^2 \sigma_0$, where $\sigma_0$ is now the single-$W$ partonic cross section, leading to the naive estimate $\sim\! 10^{-9}\, \sigma_0$. If only a factor of $\pi$ is included with each $\alpha$, one finds instead $\sim 10^{-7} \sigma_0$. The dPDF contribution $F_{u \overline d} F_{u \overline d}$ to $W^+W^+$ production is of order $\lqcd^2 \si_0^2 \sim 10^{-8} \si_0$ to $10^{-9} \si_0$.  This process has been studied in more detail in Refs.~\cite{Kulesza:1999zh,Cattaruzza:2005nu,Maina:2009sj,Gaunt:2010pi}. 

The evolution in \eq{10} vanishes in this case because $u \overline d$ cannot mix with a gluon. The same-sign $WW$ cross section also gets suppressed contributions from other flavor combinations. For example, the gluon single PDF $f_g f_g$ term contributes at order $ [\alpha/(4 \pi \sin^2\theta_W)]^2 [\alpha_s/(4 \pi)]^4 \sigma_0$ and the $F_{u \overline u} F_{d \overline d}$ dPDF term contributes at
order $ [\alpha_s/(4 \pi)]^4 \lqcd^2 \si_0^2$. Thus same-sign $WW$ production is a good process to study double parton scattering, since contamination by other production mechanisms is small.

%~~~~~~~~~~~~~~~~~~~~~~~~~~~~~~~~~~~~~~~~~~~~~~~~~~~~~~~~~~~~~~~~~~~~~~~~~~~~~~~
\section{Interpretation}
%~~~~~~~~~~~~~~~~~~~~~~~~~~~~~~~~~~~~~~~~~~~~~~~~~~~~~~~~~~~~~~~~~~~~~~~~~~~~~~~

Currently, the experimental knowledge of double parton scattering is limited to effective cross sections, which is a single number that measures the suppression of double parton scattering with respect to the corresponding single parton scattering processes. In the effective cross section approximation, the dPDF is approximated by $F(x_1,x_2,\mathbf{z}_\perp) \to F(x_1,x_2) G(\mathbf{z}_\perp) \to  f(x_1) f(x_2)G(\mathbf{z}_\perp)$ in terms of single PDFs and a universal form factor in $\mathbf{z}_\perp$. The double parton scattering rate is then given $\sigma_i \sigma_j/\sigma_{\rm eff}$, where $\sigma_{i,j}$ are the usual single parton cross-sections for the subprocesses $i,j$
and
%%%
\begin{eqnarray}
  \sieff = \bigg[\int\! \df^2 \mathbf{z_\perp}\, G(\mathbf{z_\perp})^2\bigg]^{-1}\,.
\end{eqnarray}
%%%
This approximation ignores all correlations between the partons, including in flavor, the momentum fractions and  $\mathbf{z}_\perp$.

As our knowledge of double parton scattering expands, this minimalistic way of describing the data will lead to inconsistencies and require correction. Initially, one could study flavor correlations by using $F_{ab}(x_1,x_2) \to \eta_{ab} f_a(x_1) f_b(x_1) $. Determining $uu$ vs $ud$ vs $dd$ double parton distributions allows one to study flavor correlations. Different flavor combinations can be obtained by combining various double parton processes such as  $pp \to \gamma^* \gamma^*$, $\gamma^* W$, $\gamma^* Z$, $W+\text{jets}$, etc.
In a  naive quark model, the proton has $uud$ constituents at a low scale, and one would expect that the $dd$ double parton distribution is suppressed so that $\eta_{dd}$ is smaller than $\eta_{uu}$ and $\eta_{ud}$.  
Determining the double parton distributions $F(x_1,x_2)$ which includes momentum fraction correlations could be a next step.
At some point spin correlations and correlations in $\mathbf{z_\perp}$ can no longer be ignored. In the constituent quark model, the $uu$ quark pair in a proton has probability 2/3 to have parallel spins and 1/3 to have antiparallel spins. Different processes access different spin structures (e.g.~$W^+W^+$ vs.~$ZZ$), which may help disentangling the two. There are also color correlations and interference effects, which are Sudakov suppressed~\cite{Mekhfi:1988kj,Manohar:2012jr}, and decrease rapidly with energy. In a first approximation, these can be neglected for LHC data.

We will now briefly describe how one should think about the quark double PDFs. In flavor space, the $q \overline q$ dPDFs can be decomposed into flavor singlet 
$F_S=\sum_i F_{q_i \overline q_i}/n_F$, where $n_F$ is the number of active flavors,
and non-singlet combinations. Mixing of the the quark dPDF with $q \overline q g$ operators (see Fig.~\ref{fig:DDY5}) is chirality suppressed, so they can only mix with $f_g$, which is a flavor singlet.  Only the flavor singlet combination $F_S$ mixes with $f_g$ at $\mathbf{z_\perp}=0$. 
All flavor nonsinglet combinations do not mix with $f_g$, and can be treated as separate functions $F_{NS}(x_1,x_2,\mathbf{z_\perp})$, and the product
%%%
\begin{equation}
 \int \rd^2 \mathbf{z_\perp}  F_A(x_1,x_2,\mathbf{z}_\perp)F_B(x_3,x_4,\mathbf{z}_\perp)
 \equiv \mathcal{F}_{A,B}(x_1,x_2,x_3,x_4)
\end{equation}
%%%
is the convolution of two functions where $A$ and $B$ are both non-singlets (NS).  However, when both double PDFs are singlets ($A=B=S$), the structure in \eqs{6}{10} is no longer factorizable and the integral of the two double PDFs $\mathcal{F}_{S,S}$ should be thought of as one function of four momentum fractions,
with evolution given by \eqs{10}{12} replacing Eq.~(\ref{eq:mix}). The case where $A=S$ and $B=NS$ still factorizes, but requires the singlet function to be thought of as a distribution in $\mathbf{z}_\perp$, so it may be more convenient to treat $\mathcal{F}_{S,NS}$ as a function of four variables, as for the $SS$ case.

%~~~~~~~~~~~~~~~~~~~~~~~~~~~~~~~~~~~~~~~~~~~~~~~~~~~~~~~~~~~~~~~~~~~~~~~~~~~~~~~
\section{Conclusions}
%~~~~~~~~~~~~~~~~~~~~~~~~~~~~~~~~~~~~~~~~~~~~~~~~~~~~~~~~~~~~~~~~~~~~~~~~~~~~~~~

We have studied the mixing between single and double PDFs, and shown how to solve the $\delta^{(2)}(\mathbf{z}_\perp=0)$ problem. The factorization theorem for a physical process such as double Drell-Yan has both single and double parton contributions. These mix, and so are only separately defined after a choice of renormalization scheme.  The mixing diagrams involve
$F(\mathbf{z}_\perp=0)$, which has additional divergences because the transverse separation has been set to zero. The evolution of $F(\mathbf{z}_\perp=0)$
is given in the appendix. DPS is not a physical process by itself, but one contribution to a physical process. Including all the contributions leads to a consistent  factorization theorem including QCD radiative corrections.

%~~~~~~~~~~~~~~~~~~~~~~~~~~~~~~~~~~~~~~~~~~~~~~~~~~~~~~~~~~~~~~~~~~~~~~~~~~~~~~~
\acknowledgements
%~~~~~~~~~~~~~~~~~~~~~~~~~~~~~~~~~~~~~~~~~~~~~~~~~~~~~~~~~~~~~~~~~~~~~~~~~~~~~~~

We would like to thank C.~Campagnari, F.~Golf, and A.~Yagil for introducing us to double parton scattering, and for helpful discussions.
This work is supported by DOE grant DE-FG02-90ER40546. 

\begin{appendix}
%~~~~~~~~~~~~~~~~~~~~~~~~~~~~~~~~~~~~~~~~~~~~~~~~~~~~~~~~~~~~~~~~~~~~~~~~~~~~~~~
\section{Generalized dPDFs}
%~~~~~~~~~~~~~~~~~~~~~~~~~~~~~~~~~~~~~~~~~~~~~~~~~~~~~~~~~~~~~~~~~~~~~~~~~~~~~~~

There is also a subtlety at $\mathbf{z}_\perp = 0$ for the non-mixing contributions to the RGE. First of all, it should be noted that at $\mathbf{z}_\perp = 0$ the interference dPDFs can be related to the standard dPDFs, because the fields are no longer separated in $\mathbf{z}_\perp$. Also the spin structures involving an explicit $\mathbf{z}_\perp$ vanish. As discussed in Refs.~\cite{Diehl:2011tt,Diehl:2011yj,Manohar:2012jr}, the effect of soft radiation is nontrivial, and is described by a soft function depending on $\mathbf{z}_\perp\sim 1/\lqcd$. The soft function disappears at $\mathbf{z}_\perp=0$, since $S(\mathbf{z}_\perp=0)=1$, implying a change in the dPDF anomalous dimensions at $\mathbf{z}_\perp = 0$. There are no rapidity divergences at $\mathbf{z}_\perp=0$, so the $\nu$ anomalous dimension vanishes~\cite{Chiu:2011qc,Chiu:2012ir}.

Define the generalized $F_{q\bq}^T$ dPDF distributions by
\begin{widetext}
%%%
\begin{eqnarray}
 && \delta (p^- - p^{\prime-}+\omega_1+\omega_2-\omega_3-\omega_4)
F_{q\bq}^T(\omega_1,\omega_2,\omega_3,\omega_4,\mathbf{z}_\perp) 
=  \int \frac{\rd z_1^+}{4\pi} \frac{\rd z_2^+}{4\pi} \frac{\rd z_3^+}{4\pi} \frac{\rd z_4^+}{4\pi}\ e^{i( \omega_3 z_3^+
+ \omega_4 z_4^+ -\omega_1 z_1^+  -  \omega_2 z_2^+)/2}  \nn
&&    
(-4\pi)\, \sqrt{p^- p^{\prime -}}  \braket{p^\prime| \overline T \left\{ \Big[\overline \psi (z_1^+,0,\mathbf{z}_\perp)\frac{\bnslash}{2} T^A \Big]_a   \psi_b (z_2^+,0,\mathbf{0}_\perp) \right\} T \left\{ \psi_a (z_3^+,0,\mathbf{z}_\perp) \Big[\overline \psi (z_4^+,0,\mathbf{0}_\perp)\frac{\bnslash}{2} T^A \Big]_b\right\}  | p} \,,
\label{53}
\end{eqnarray}
\end{widetext}
%%%
with analogous definitions for the other dPDFs. (We have suppressed the Wilson lines, which are necessary for gauge invariance, see e.g.~Ref.~\cite{Manohar:2012jr}.) The dPDFs for DPS are then
%%%
\begin{eqnarray}
F_{q\bq}^T(x_1=\omega_1/p^-,x_2=\omega_2/p^-,\mathbf{z}_\perp)  &=& F_{q\bq}^T(\omega_1,\omega_2,\omega_1,\omega_2,\mathbf{z}_\perp)
\,,\nn\label{54}
\end{eqnarray}
%%%
with the momentum fractions equal in pairs, and $p=p^\prime$.

The RGE evolution of the generalized dPDFs at $\mathbf{z}_\perp \not=0$ maintains the equality $\omega_1=\omega_3$ and $\omega_2=\omega_4$, and can be  obtained from the results given in Ref.~\cite{Manohar:2012jr}. At $\mathbf{z}_\perp=0$, the divergence structure changes. In the forward direction $p=p^\prime$, the generalized dPDF in Eq.~(\ref{53}) is a function $F_{q\bq}(\omega_1,\omega_2,\omega_3,\omega_1+\omega_2-\omega_3,\mathbf{0}_\perp)$ of three variables, and the RGE evolution mixes all three variables, and does not preserve the relations $\omega_1=\omega_3$, $\omega_2=\omega_4$. $F_{q\bq}(\omega_1,\omega_2,\omega_3,\omega_1+\omega_2-\omega_3,\mathbf{0}_\perp)$  is the light cone correlator of four fields. The interference dPDFs at $\mathbf{z}_\perp=0$ are given by $I_{q\bq}^T(\omega_1,\omega_2,\omega_3,\omega_4,\mathbf{0}_\perp)=F_{q\bq}^T(\omega_2,\omega_1,\omega_3,\omega_4,\mathbf{0}_\perp)$, and similarly for the other color and spin structures.

We give here the anomalous dimensions for the generalized dPDF in Eq.~(\ref{53}) at $\mathbf{z}_\perp=0$, from which  $p=p^\prime$ follows as a special case. 
The evolution equation is
\begin{widetext}
%%%
\begin{eqnarray}
\mu \frac{\rd}{\rd \mu} F(\omega_1,\omega_2,\omega_3,\omega_4) &=& \frac{\alpha_s}{\pi} \int \rd \omega_1^\prime \rd \omega_2^\prime \rd \omega_3^\prime \rd \omega_4^\prime\
P(\omega_1,\omega_2,\omega_3,\omega_4;  \omega_1^\prime , \omega_2^\prime , \omega_3^\prime,  \omega_4^\prime)\
F( \omega_1^\prime,  \omega_2^\prime,  \omega_3^\prime,  \omega_4^\prime)
\end{eqnarray}
%%%
and the evolution kernel is
%%%
\begin{eqnarray}
P&=& \left[ \begin{array}{cc}
C_F & 0 \\ 0 & C_F - \frac12 C_A
\end{array} \right] \left( \Pdef_{1324}^{(\zeta)}+\Pdef_{2413}^{(\zeta)}\right) -\left[ \begin{array}{cc}
C_F & 0 \\ 0 & C_F
\end{array} \right]\delta(\omega_1^\prime-\omega_1)\delta(\omega_2^\prime-\omega_2) \delta(\omega_3^\prime-\omega_3)
 \delta(\omega_4^\prime-\omega_4) \\[5pt]
&&+\left[ \begin{array}{cc}
0 & 0 \\ 0 & \frac14 C_A
\end{array} \right] \left[ \Pdef_{2314}^{(X/2)}+\Pdef_{1423}^{(X/2)} +\widetilde \Pdef_{1234}+\widetilde \Pdef_{3412} \right]\pm\left[ \begin{array}{cc}
0 & 1 \\ C_1 & \frac14 C_d 
\end{array} \right] \left[ \Pdef_{2314}^{(X/2)}+\Pdef_{1423}^{(X/2)}- \widetilde \Pdef_{1234}-\widetilde \Pdef_{3412}\right] ,\nn[10pt]
\Pdef_{abcd}^{(M)} &=& \delta(\omega_a^\prime  - \omega_b^\prime+ \omega_b -\omega_a)\delta(\omega_c^\prime-\omega_c)
\delta(\omega_d^\prime-\omega_d)
\left[\frac{1}{\omega_a^\prime}   \left( \frac{\omega_a}{\omega_a^\prime-\omega_a} \right)_+  + \frac{1}{\omega_b^\prime}  \left( \frac{\omega_b}{\omega_b^\prime-\omega_b} \right)_+  +  \left( \frac{\omega_a^\prime-\omega_a}{\omega^\prime_a \omega^\prime_b} \right) \theta(\omega_a^\prime - \omega_a) M \right], \nn
\widetilde \Pdef_{abcd} &=& \delta(\omega_a^\prime  + \omega_b^\prime - \omega_a -\omega_b)\delta(\omega_c^\prime-\omega_c)
\delta(\omega_d^\prime-\omega_d)
\left[\frac{1}{\omega_a^\prime}   \left( \frac{\omega_a}{\omega_a^\prime-\omega_a} \right)_+  + \frac{1}{\omega_b^\prime}  \left( \frac{\omega_b}{\omega_b^\prime-\omega_b} \right)_+ + \frac12 X \begin{cases}
\frac{\omega_a}{\omega_a^\prime(\omega_a^\prime + \omega_b^\prime)} & \omega_a \le \omega_a^\prime    \\
\frac{\omega_b}{\omega_b^\prime(\omega_a^\prime + \omega_b^\prime)} & \omega_b \le \omega_b^\prime    \\
\end{cases}
 \right], \nonumber
\label{58}
\end{eqnarray}
%%%
with $\pm$ for the $qq$ and $q \overline q$ cases. 
The plus distributions are 
%%%
\begin{eqnarray}
 \left( \frac{\omega}{\omega^\prime-\omega} \right)_+ &\equiv& \lim_{\beta \to 0} \quad \frac{\omega}{\omega^\prime-\omega}\  \theta(\omega^\prime-\omega-\beta)
 + \delta(\omega^\prime-\omega-\beta)\left(1-\frac{\beta}{\omega^\prime} + \ln \frac{\beta}{\omega^\prime}\right)\,.
\end{eqnarray}
%%%
\end{widetext}

The color factors are $C_1=(N^2-1)/(4N^2)=2/9$, $C_d=(N^2-4)/N=5/3$. The matrices shown explicitly in Eq.~(\ref{58}) are in $(F^1,F^T)$ color space, and $M=X,\zeta$, where
%%%
\begin{eqnarray}
X = \left(\begin{array}{ccc}
1 & -1 & 0 \\ 
-1 & 1 & 0  \\
0 & 0 & 2 
\end{array}\right),\qquad
\zeta = \left(\begin{array}{ccc}
1 & 0 & 0 \\ 
0 & 1 & 0  \\
0 & 0 & 0
\end{array}\right)
\end{eqnarray}
%%%
are  matrices in $qq$, $\Delta q \Delta q$, $\delta q \delta q$ spin space. $X$ is the restriction of the matrix $X$ in Ref.~\cite{Manohar:2012jr} to these three spin structures. The other spin structures vanish at $\mathbf{z}_\perp=0$. 

\end{appendix}

\newpage

\bibliography{dps}
 
\end{document}